\documentclass[%
 reprint,
 amsmath,amssymb,
 aps,superscriptaddress]{revtex4-1}
\usepackage{graphicx}
\usepackage{dcolumn}
\usepackage{bm}
\begin{document}

\title{Response Properties of Axion Insulators and Weyl Semimetals Driven by Screw Dislocations and Dynamical Axion Strings}

\author{Yizhi You}
\affiliation{Department of Physics, University of Illinois, 1110 W. Green St., Urbana, Illinois 61801-3080, USA}
\author{Gil Young Cho}
\affiliation{Department of Physics, Korea Advanced Institute of Science and Technology, Daejeon 305-701, Korea}
\author{Taylor L. Hughes}
\affiliation{Department of Physics, University of Illinois, 1110 W. Green St., Urbana, Illinois 61801-3080, USA}

\date{\today}
\begin{abstract}
In this paper, we investigate the theory of dynamical axion strings emerging from chiral symmetry breaking in three-dimensional Weyl semimetals. The chiral symmetry is spontaneously broken by a charge density wave (CDW) order which opens an energy gap and converts the Weyl semimetal into an axion insulator. Indeed,  the phase fluctuations of the CDW order parameter act as a dynamical axion field $\theta({\vec{x}},t)$ and couple to electromagnetic field via $\mathcal{L}_{\theta}=\frac{\theta(\vec{x},t)}{32\pi^2} \epsilon^{\sigma\tau\nu\mu} F_{\sigma\tau} F_{\nu\mu}.$ Additionally, when the axion insulator is coupled to deformations of the background geometry/strain fields via torsional defects, e.g., screw dislocations, there is a novel interplay between the crystal dislocations and dynamical axion strings.  For example, the screw dislocation traps axial charge, and there is a Berry phase accumulation when an axion string (which carries axial flux) is braided with a screw dislocation. In addition, a cubic coupling between the axial current and the geometry fields is non-vanishing and indicates a Berry phase accumulation during a particular three-loop braiding procedure where a dislocation loop is braided with another dislocation and they are both threaded by an axion string. We also observe a chiral magnetic effect induced by a screw dislocation density in the absence of a nodal energy imbalance between Weyl points, and describe an additional chiral geometric effect and a geometric Witten effect.

\end{abstract}

\maketitle

\section{Introduction and motivation}

Classifying states of matter is one of the primary goals of condensed matter physics. First, the study of symmetries led to the Landau classification of ordered states of matter that spontaneously break a subset of the microscopic symmetries. Decades later it was discovered that even systems with no broken symmetries may still be distinguished by their topological order\cite{levin2006detecting,wen1990topological}. More recently, the combination of symmetry and topology have led to exciting developments in the field of symmetry-protected topological phases, such as time-reversal invariant topological insulators\cite{kane2005z,schnyder2008classification,kitaev2009periodic,qi2008topological,chen2010local,metlitski2013bosonic,wang2014classification,metlitski2014interaction,chen2012symmetry,vishwanath2013physics}.While the underlying mathematics allows one to distinguish these phases in principle, often it is important how to distinguish them in practice in an experiment. Helpfully, some of the most striking effects of conventionally ordered, and topologically ordered, phases can be seen in their electromagnetic properties. For example, a ferromagnet has a spontaneously generated magnetic moment, and the fractional quantum Hall effect has a quantized fractional Hall conductance.

While the electromagnetic response properties provide remarkable signatures, they are not the full story. For example, in topological superconductors there will be no such quantized electromagnetic response coefficients. In these cases one can resort to studying the response to strain, thermal gradients,  or, in general, deformations of the geometry\cite{ryu2012electromagnetic,wang2011topological,wen1992shift,wen1988spin}. Superconductors are not the only systems that are sensitive to these probes, as recently there has been an intense research effort on the geometric response of integer/fractional quantum Hall systems\cite{avron1995,you2014theory,gromov2015framing,you2016nematic,cho2014geometry,bradlyn2012kubo,read2009non,hoyos2012hall}, topological insulators\cite{hughes2013torsional,you2013field,hughes2011torsional,you2013field}, and topological semimetals\cite{parrikar2014torsion,zhang2015topological,hughes2011torsional,read2011hall}. These systems all have interesting electromagnetic properties, but their geometric responses, for example the Hall viscosity\cite{avron1995,hoyos2012hall}, provide a more complete picture. 

The majority of the work on geometric response has been directed at two dimensional systems, but in this article we focus on 3D systems of interest: the spontaneously-generated axion insulator and its parent state the Weyl semimetal (WSM)\cite{wang_zhang}. The axion insulator studied here is a 3D gapped phase with a dynamical axion $\theta$-angle. It arises from a WSM state that is unstable to the formation of a charge density wave (CDW). Once the CDW forms, the WSM is gapped, and the resulting insulating state has a dynamical ``axion" field $\theta(\vec{x},t)$: the phase of the CDW order parameter. Both the parent WSM state and the axion insulator (AI) have unusual electromagnetic and geometric responses\cite{ramamurthy2014patterns,parrikar2014torsion,chen2013axion,zyuzin2012topological,wan2011topological,burkov2011weyl,liu2013chiral,wang2016topological,cho2011possible}, and we predict even more observable phenomena in this article, including responses that rely on an interplay between the conventional CDW order parameter and the topological electronic structure. 

A main focus of this article is to study the interplay between the dynamical axion string (i.e., a vortex in the CDW order parameter) and the background torsional defects arising from crystal lattice dislocations. We demonstrate that the effective coupling between axion and torsion fields indicates a (``two-loop") statistical string interaction between an axion string and a screw dislocation line, as well as a three loop statistical interaction between two dislocation loops threaded by an axion string. Most importantly, we show that this system can also exhibit a dislocation driven chiral magnetic effect, i.e., an electromagnetic response created from a geometric origin. 

The main result of this work is the effective electromagnetic and geometric theory for the anomalous response of a dynamical axion insulator coupled to geometric deformations. We find several new effects due to the coupling between a dynamical axion field (e.g., axion string defects) and lattice dislocations including new examples of purely geometric responses as well as mixed geometric-electromagnetic phenomena. For example, we show an electromagnetic response effect driven by lattice dislocation density. Our article is organized as follows. In Section II we review the instability of a WSM to form an AI phase in the presence of CDW order. We also provide the theoretical setup for the inclusion of axion string defects and geometric deformations. In Section III we review the anomalous response of WSMs to electromagnetic and geometric perturbations. We review the axial and stress anomalies, including a geometric contribution to the axial anomaly, and show how the theory depends on the momentum/energy separation of the Weyl nodes. Section IV presents new results focused on the interplay between axion strings and lattice dislocations in the AI system; we mention a mutual statistical interaction and a possible three-loop statistical interaction. In Section V we present possible electromagnetic response properties of the WSM and AI phases due to the dynamical axion field and the existence of lattice dislocations. We also talk about two types of analogous geometric effects including a geometric version of the chiral magnetic effect and a geometric Witten effect.  We conclude in Section VI and give some calculational details in Appendix A.

\section{Axion Insulator and Dynamical Axion Strings from Charge Density Wave Fluctuations}
\subsection{Axion Insulator Instability in a Weyl Semimetal}
Let us begin by briefly reviewing the theory of the dynamical axion insulator as proposed in Ref. \onlinecite{wang_zhang}, and then discussing how this theory is coupled to geometric perturbations. The AI is generated by spontaneous chiral symmetry breaking in three dimensional Weyl semimetals\cite{wang_zhang,bi2015unidirectional}. For a (two-node) WSM, breaking the chiral symmetry is accomplished by a commensurate translation symmetry breaking that nests the Weyl nodes. By turning on an attractive interaction to induce a CDW instability that nests two Weyl points, the Weyl cones can be gapped, and the resulting gapped state is an  �axion insulator.�  Hence, the spontaneous chiral symmetry breaking order parameter is just a CDW order parameter, and the phase of the CDW ($\theta(\vec{x},t)$), i.e., the Goldstone mode, is identified as the axion field. 

To proceed more explicitly, we can start from a simple WSM with only two nodes, and gap out the two Weyl cones by spontaneous chiral symmetry breaking. The Weyl semimetal is stable in the presence of weak interactions if we prohibit pairing and scattering between the two Weyl cones. A four-fermion interaction with a finite coupling constant $g$ can induce chiral symmetry breaking (along with translation symmetry breaking). This symmetry breaking dynamically generates an energy gap with magnitude $2\vert m\vert$ that lowers the ground state energy compared to that of a nominally gapless system. Indeed, Ref. \onlinecite{wang_zhang} showed that when a critical interaction strength is reached, an effective mean-field Hamiltonian with broken chiral symmetry written as
\begin{eqnarray}
&&\mathcal{H}=\nonumber\\ && \Psi^{\dagger}(\bm{r},t) (p_i \tau_3 \otimes \sigma_i+b_{i}\sigma_{i}+m_1\tau_1+ m_2 \tau_2)\Psi(\bm{r},t)-\frac{ |\vec{m}|^2 }{g}\nonumber\\
&&\vec{m}\to m_1+im_2=|m|e^{i\theta},\theta=-2\bm{b}\cdot \bm{r}
\label{Hwsm}
\end{eqnarray}\noindent is expected.
 The first two terms are the effective Hamiltonian of the WSM where the $\sigma^{i}$ are spin Pauli matrices, the $\tau^{i}$ are Pauli matrices for the valley index, and $i=1,2,3.$ The shift vector $b_{i}$ is half of the momentum separation between two Weyl cones/valleys. The mass term should be identified as a CDW order parameter which nests between the two Weyl cones,
and $2\bm{b}$ is the scattering wave vector  between the two Weyl points. 

Interestingly, in the gapped phase, the WSM is converted into an axion insulator. The phase angle field $\theta$ of the CDW vector $\vec{m}$ is a Goldstone mode that couples with the electromagnetic response in the form of a theta term\cite{wang_zhang}:
\begin{eqnarray}
\mathcal{L}_{\theta}=\frac{\theta(\vec{x},t)}{32\pi^2} \epsilon^{\sigma\tau\nu\mu} F_{\sigma\tau} F_{\nu\mu}.
\end{eqnarray}
Additionally,  if there is a vortex-like configuration in the CDW order, the defect line acts as a dynamical axion (or cosmic) string which carries gapless chiral modes, where the chirality is determined by the axion string vorticity. It has also been shown that it is natural to view the axion strings in this model as dislocations of the CDW crystal order\cite{wang_zhang}. Thus, since our interest is in the geometric response of such materials, we might expect an interesting interplay between the response due to deformations of the underlying lattice and that of the electronic CDW order. 

\subsection{Coupling to Background Fields}

We are now ready to set the stage for our work by first reviewing the properties of an axion string configuration, and then coupling this model to the background geometry of the distorted underlying lattice.
Let us assume we are in the gapped AI phase, and that a vortex (axion string) of the CDW phase angle (i.e., a dislocation of the CDW order) is present such that
\begin{eqnarray}
\vec{m}\to m_1+im_2= |m|e^{i\theta},~\theta= \theta_v(\vec{x},t)-2\bm{b}\cdot \bm{r}
\label{H}
\end{eqnarray}\noindent where $\theta_v(\vec{x},t)$ is the vortex field.
In the presence of such a configuration, one can always eliminate the spatial dependence of the mass term by performing the chiral transformation
\begin{align}
 &R^{\dagger}(\vec{x})(m_1(\vec{x})\tau_1+m_2(\vec{x})\tau_2)R(\vec{x})=\vert m\vert \tau_1,\nonumber\\
 &\Psi'(\vec{x})=R(\vec{x}) \Psi(\vec{x})\nonumber\\
 &R(\vec{x})=e^{-i\frac{\theta(\vec{x})}{2}\tau_3}.
\end{align}
After the chiral gauge transformation, the mass term becomes vortex free, but there is a side-effect because  a new axial gauge field term is introduced in the Lagrangian:
\begin{eqnarray}
&\mathcal{L}= \bar \Psi  (p_0 \tau_1+B_0 \tau_2+p_i \tau_2 \otimes \sigma_i+B_i \tau_1 \otimes \sigma_i+m)\Psi \nonumber\\
& B_{\mu}=\partial_{\mu}\theta_v/2=\epsilon^{ij}\frac{m_i \partial_{\mu} m_j}{2 \vert  m\vert^2},\label{eq:bfromvortex}
\end{eqnarray}\noindent where $\bar \Psi =-i\Psi^{\dagger} \tau_1$ and we have dropped the primes on the fermion fields for convenience. Note that the chiral transformation removed the term proportional to $\vec{b}$ in the Lagrangian that would have otherwise contributed to $B_{\mu}$ as a constant shift.  Apart from the change of the Lagrangian, it is well known that such a chiral transformation also generates an anomalous term from the fermion path integral measure $\mathcal{S}_{anomalous} \sim \theta(x) F \wedge F,$ which we will come back to later\cite{fujikawa1980comment,fujikawa1984evaluation}. 

Now let us apply a strain/displacement of the background lattice. This is introduced through a distorted metric tensor, which we decompose into the usual frame-fields. Thus, we introduce the spatial frame fields $e^{a}_{i},$ $a=x,y,z$ and $i=x,y,z$\footnote{These are really the co-frame fields, but we will not bother to distinguish and abuse the language to call these frame fields as well.} and couple them to our theory. For the majority of our article we will only focus on a specific set of distortions in order to represent our main point of interest: screw dislocations oriented parallel to the $z$-direction (i.e., parallel to the axion string for convenience). The relevant distortions are contained in the $e^{z}_x$ and $e^{z}_{y}$ fields which, in terms of the usual elasticity fields,  are written
\begin{eqnarray}
e^z_x=\partial_x u^z,~e^z_y=\partial_y u^z,
\end{eqnarray} where $u^z$ is the lattice displacement in the $z$-direction. The diagonal elements of the frame are undistorted $e_{x}^{x}=e_{y}^{y}=e_{z}^{z}=1,$ and all other components are chosen to vanish.
 Hence, if $u^z$ is non-uniform, then the background metric is distorted, and the distortion is reflected by the non-vanishing frame fields $(e^z_x,~e^z_y)$.  If we consider a screw dislocation parallel to the $z$-axis, then traveling around the dislocation line we are expected to accumulate a Burgers translation vector with a $z$-component: ${\cal{B}}^z=\oint e^z_i dx^i.$

The final model in the presence of an axion string configuration and the geometric distortion is written as
\begin{align}
\mathcal{L}= &\bar \Psi (p_0 \tau_1+B_0 \tau_2+p_i \tau_2 \otimes \sigma_i+B_i \tau_1 \otimes \sigma_i+m \nonumber\\
&+e^z_x p_z \tau_2 \otimes \sigma_x+e^z_y  p_z \tau_2 \otimes \sigma_y)\Psi.
\end{align}
We have seen that the $B_\mu$ field originates from a vortex of the mass term, i.e., a dislocation of the CDW order. Thus, as mentioned above, it is natural to expect some interesting interplay between the axion string appearing from the CDW defect and a screw dislocation of the underlying crystal lattice. 
Remarkably, as we discuss further below, after introducing the frame fields, the chiral transformation we performed to remove the phase of the mass term will  generate additional anomalous terms  from the path integral measure that depend on the geometric distortion. Our goal in the remainder of the article is to carefully derive and discuss these anomalous response terms.

\section{Geometric Response in a Weyl Semimetal}
\subsection{Review of the Electromagnetic Response from the Axial Anomaly}
Before we detail the properties of the AI phase, we first discuss the effective theory of the interplay between axion strings and crystal dislocation lines in Weyl semimetals, i.e., the gapless parent state of the AI.
We start from the WSM phase with two Weyl nodes (with nodal separation $2b_i$), and turn on the electromagnetic field ($A_\mu$) and an axial gauge field ($A^{ax}_\mu$):
\begin{eqnarray}
&&\mathcal{L}= \bar \Psi (D_0 \tau_1+A^{ax}_0 \tau_2+D_i \tau_2 \otimes \sigma_i+A^{ax}_i \tau_1 \otimes \sigma_i)\Psi\nonumber\\
&&D_{\mu} =p_{\mu}+A_{\mu}.
\end{eqnarray}
Note that since both the momentum separation $2b_i$ and $A^{ax}_{\mu}$ couple to the axial current in the same way, we can always absorb $\vec{b}$ into the axial gauge field.

For a WSM, even though the axial current is conserved at the classical level, it is well known that it exhibits a quantum axial anomaly: 
\begin{eqnarray}
\partial_{\mu} J^{ax}_{\mu}=\frac{1}{4\pi^2} \epsilon^{\mu \nu \rho \lambda} \partial_{\mu} A_{\nu}  \partial_{\rho} A_{\lambda}.
\end{eqnarray}
As a result, the leading effective coupling between the axial gauge field and external electromagnetic field is:
\begin{eqnarray}
\mathcal{L}=\frac{1}{4\pi^2} \epsilon^{\mu \nu \rho \lambda} A^{ax}_{\mu} A_{\nu}  \partial_{\rho} A_{\lambda}.
\end{eqnarray} This coupling is known to give rise to many interesting phenomena\cite{nielsen1983,burkov2014anomalous,ran2009one,chen2013axion,zyuzin2012topological,wan2011topological,burkov2011weyl}.
First, if  there is a momentum separation $2b_i$ between two Weyl cones at low energy, this acts like a background contribution to the axial gauge field, i.e., $A_{i}^{ax}= b_i$, and the WSM exhibits a (non-quantized) anomalous Hall effect. One heuristic way to understand this phenomenon is to treat the Hamiltonian for each $k_z$ as a gapped 2D insulator (except for the two nodal points). Hence the theory is equivalent to many copies of a 2D massive Dirac fermion (i.e., a Chern insulator) at each value of $k_z,$ and each value of $k_z$ will have a quantized Hall conductance. The $k_z$ values between the nodes contribute a value with a magnitude of $e^2/h,$ while those outside the nodes contribute a vanishing amount. Thus, in total  we expect a finite Hall conductance coming from the sum over all  $k_z$ values between the nodes.  
On the other hand, if we turn on an energy difference $2b_0$ between the two Weyl nodes, one naively expects a chiral magnetic effect due to the contribution:
\begin{eqnarray}
\mathcal{L}=\frac{1}{4\pi^2} \epsilon^{0 \nu \rho \lambda} b_{0} A_{\nu}  \partial_{\rho} A_{\lambda}.
\end{eqnarray}

\subsection{Path Integral Measure for a Chiral Transformation with Distorted Geometry}

The anomalous Hall effect and the chiral magnetic effect indicate time-reversal breaking responses in the plane perpendicular to the momentum separation between the Weyl cones (at least when there are only two nodes). These are both a consequence of the axial anomaly.  Usually a Hall effect is accompanied by a Hall viscosity\cite{avron1995} and hence, one also expects a similar Hall viscosity response associated to a stress-tensor anomaly in a Weyl semimetal\cite{parrikar2014torsion,sun2014chiral}. Indeed, even though the stress-energy tensor is  conserved at the classical level, one can verify that there is an anomaly
\begin{eqnarray}
\partial_{\mu} T^{a}_{\mu} =\frac{\Lambda^2}{2\pi^2} \epsilon^{\rho \lambda \nu \mu}  \partial_{\lambda} A^{ax}_{\rho}  \partial_{\nu} e^{a}_{\mu} 
\end{eqnarray}\noindent where $\Lambda$ is a high-energy cut-off.
Thus, we see that the stress-energy tensor is anomalous, and the non-vanishing contribution implies a coupling between the axial gauge field (including the momentum/energy differences in the Weyl nodes) and the frame field. As an aside we note that this is reminiscent of a mixed ``$E\cdot B$" term if we define the axial and ``geometrical" field-strengths as\footnote{The definition of the torsional geometrical field-strength is imprecise since it should include contributions from the spin connection. However, in this article our focus is on torsion instead of curvature so we have chosen a gauge where the spin-connection is vanishing.}  \begin{eqnarray}
F_{\mu\nu}^{(ax)}&\equiv&\partial_{\mu} A^{ax}_{\nu}-\partial_{\nu} A^{ax}_{\mu}\nonumber\\
 F_{\mu\nu}^{(geom),a} &\equiv& \partial_{\mu} e^{a}_{\nu}-\partial_{\nu} e^a_{\mu}.\nonumber
\end{eqnarray}
Using these definitions,  the stress anomaly can be expressed as: 
\begin{eqnarray}
\partial_{\mu} T^{a}_{\mu} =\frac{\Lambda^2}{2\pi^2}(\vec{E}^{(ax)} \cdot \vec{B}^{(geom),a}+\vec{E}^{(geom),a} \cdot \vec{B}^{(ax)}).\nonumber\\
\end{eqnarray}  
 The physical interpretation of the anomaly is that if we have an axial magnetic flux $B^{(ax)}_i$, i.e., an axion string, along the $i$-direction it will trap a chiral fermion, and hence a non-vanishing momentum density. Subsequently, applying an $\vec{E}^{(geom),a}$ (e.g., a time dependent strain/stretching) will lead to an anomalous momentum current due to the geometric deformation of the chiral fermion states. There is an additional phenomena associated to this anomaly as well: if we have a screw dislocation along the $i$-direction with a Burgers vector $B^{(geom),a}={\cal{B}}^a$, then the application of an axial electric field will also drive an anomalous momentum current. Unfortunately, it is not immediately clear how to relate this anomalous current to the perturbation of microscopic degrees of freedom localized on the dislocation. 

In addition to a stress anomaly there is an additional geometric contribution to the usual axial anomaly (due to the Nieh-Yan term\cite{nieh1982identity}). If we consider the coupling between the axial gauge field and the frame fields, then the axial anomaly gains an additional term and should be modified as: 
\begin{eqnarray}
&&\partial_{\mu} J^{ax}_{\mu}=\frac{1}{2\pi^2}(\vec{E} \cdot \vec{B}+\Lambda^2 \vec{E}^{(geom),a} \cdot \vec{B}^{(geom),a}).\nonumber\\
\end{eqnarray} The phenomena associated to the geometric contributions to this anomaly are discussed in detail in Ref. \onlinecite{parrikar2014torsion}.

In an experimental context, we expect that these anomalies might play a role in the physics of WSMs since a WSM has an effective background axial gauge field due to the momentum and energy separation of the Weyl nodes. To calculate the effects of these anomalies for our simple WSM continuum model, we can couple our theory to a background geometry as before:
\begin{eqnarray}
&&\mathcal{L}= \bar \Psi (p_0 \tau_1+p_i \tau_2 \otimes \sigma_i+b_i \tau_1 \otimes \sigma_i +e^i_j p_i \tau_2 \otimes \sigma_j)\Psi,\nonumber\\
\end{eqnarray}\noindent and then calculate the change in the path-integral measure during a chiral transformation. 
We can perform a chiral transformation to the Weyl fermion:
\begin{eqnarray}
R(\vec{x})=e^{i b_\mu x_\mu \tau_3}, ~\Psi'(\vec{x})=R(\vec{x}) \Psi(\vec{x}),
\end{eqnarray}
such that the momentum/energy separation proportional to $b_\mu$ in the effective action is eliminated in the Lagrangian. However, the chiral transformation of the Weyl fermion fields results in a change of the path integral measure which compensates for the missing term in the Lagrangian. Following the Fujikawa method\cite{fujikawa1979path,sun2014chiral}, the anomalous term from the path integral measure is
\begin{align}
\mathcal{S}_{anomaly}=\int d^4 x\Big[ &\frac{b_\sigma x_\sigma}{4\pi^2} \epsilon^{\mu \nu \rho \lambda} \partial_{\mu} A_{\nu}  \partial_{\rho} A_{\lambda}\nonumber\\
&+\frac{\Lambda^2 b_\sigma x_\sigma}{4\pi^2} \epsilon^{\mu \nu \rho \lambda} \partial_{\mu} e^l_{\nu}  \partial_{\rho} e^l_{\lambda}\Big].
\end{align}
The response equations derived from the anomalous effective action represent the chiral magnetic effect, the anomalous Hall effect, and the Hall viscosity of the WSM, and they all depend on the momentum/energy separation of the nodes through $b_{\mu}.$ We note in passing that the path integral measure change we derived here with respect to chiral transformations also applies to a massive Dirac fermion, e.g., the AI phase of a gapped WSM.

Before moving to the next sectio,n let us briefly look at the leading response to the background geometry. We find:
\begin{eqnarray}
&&\mathcal{L}= \frac{\Lambda^2}{4\pi^2} \epsilon^{\mu \nu \rho \lambda} b_{\mu} e^l_{\nu}  \partial_{\rho} e^l_{\lambda}
\label{oddviscosity}
\end{eqnarray}\noindent which was also studied in Refs. \onlinecite{hughes2013torsional,parrikar2014torsion}.
Let us look at a particular contribution to this term where the first frame field factor $e^l_{\nu}=e^i_{i}=1$ in Eq. (\ref{oddviscosity}), from which we find $\frac{\Lambda^2}{4\pi^2} \epsilon^{\mu\rho \nu \lambda} b_{\mu} \partial_{\rho} e^{\nu}_{\lambda}.$ Leading into the next section where there is a fluctuating axion field, we see that this term hints at a coupling between  axion strings/flux and lattice screw dislocations via a term of the form $\frac{\Lambda^2}{4\pi^2} \epsilon^{\mu\rho \nu \lambda} B_{\mu} \partial_{\rho} e^{\nu}_{\lambda}$ where $B_\mu$ contains the axion string configuration. As we will discuss in the next section, in the AI phase, one physical consequence of this term is to generate a kind of statistical interaction when an axion string is dragged around a lattice screw dislocation line.

\section{Interplay Between Axion String and Lattice defects in Axion Insulators}
Now let us consider the axion insulator phase generated from a WSM with CDW order. In this phase the Weyl fermions have developed a mass from the Weyl node nesting. Let us consider a case where the mass term has a vortex/axion string configuration:
\begin{eqnarray}
\mathcal{L}= &&\bar \Psi (p_0 \tau_1+p_i \tau_2 \otimes \sigma_i+m \cos(\theta_v) \tau_1+ i m \sin(\theta_v)  \tau_2,\nonumber\\
&&+e^z_x p_z \tau_2 \otimes \sigma_x+e^z_y  p_z \tau_2 \otimes \sigma_y)\Psi.
\end{eqnarray}
Note that  in this section we will focus only on the phenomena associated to the axion string configuration, and not the background term coming from $-2{\bf{b}}\cdot{\bf{r}}.$ Contributions from $b_i$ will enter the full response, but they will not be our focus for now, as some of them have been discussed perviously\cite{hughes2013torsional,parrikar2014torsion}. After we perform a chiral transformation $R(\vec{x})=\exp\left(i\frac{\theta_v(\vec{x})}{2}\tau_3\right)$, we can remove phase of the axial mass by introducing the axial gauge field configuration $B_{\mu}$ (c.f. Eq. \ref{eq:bfromvortex}) coupled with the axial current. Hence, we have the Lagrangian
\begin{eqnarray}
\mathcal{L}= &&\bar \Psi (p_0 \tau_1+B_0 \tau_2+p_i \tau_2 \otimes \sigma_i+B_i \tau_1 \otimes \sigma_i+m,\nonumber\\
&&+e^z_x p_z \tau_2 \otimes \sigma_x+e^z_y  p_z \tau_2 \otimes \sigma_y)\Psi.\label{eq:LagrangianAIB}
\end{eqnarray}

Beyond the change in the Lagrangian, the chiral transformation will also change the path integral measure. Thus, the effective theory describing the interplay between the CDW axion string defect and the lattice dislocation defect  consists of two parts: (i) $\mathcal{S}_{L}$ is the action we obtain from integrating out the gapped fermions, and (ii) $\mathcal{S}_{anomaly}$ is the action coming from the contribution of the path integral measure.
By integrating out the fermions in the Lagrangian in Eq. \ref{eq:LagrangianAIB} and performing the loop expansion to quadratic order (see Fig. \ref{fig:ebloop}), we first find:
\begin{figure} [h!] 
\begin{center}  
\includegraphics[height=1.1in,width=2.2in]{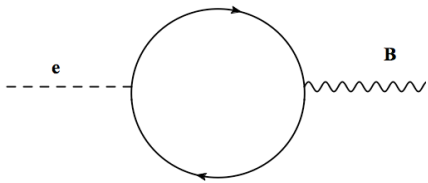}  
\caption{Coupling between the frame field and axial gauge field.}\label{fig:ebloop}
\end{center}  
\end{figure}  
\begin{eqnarray}
&\mathcal{S}_{L}=\frac{1}{4\pi^2} (\Lambda^2-m^2 \text{ln} (\frac{\Lambda}{m}))\int d^4 x \epsilon^{z\mu\nu\rho} B_{\mu} \partial_{\nu} e^{z}_{\rho}.
\label{axion_dislocation}
\end{eqnarray}
The chiral transformation we made generates an extra term from path integral measure given by\cite{sun2014chiral},
\begin{eqnarray}
&\mathcal{S}_{anomaly}=-\frac{\Lambda^2}{4\pi^2} \int d^4 x  \epsilon^{z\mu\nu\rho} B_{\mu} \partial_{\nu} e^{z}_{\rho}.
\end{eqnarray}
Thus, the total effective theory is the sum of these two terms:
\begin{eqnarray}
&\mathcal{S}_{1}=-\frac{1}{4\pi^2} (m^2 \text{ln} (\frac{\Lambda}{m}))\int d^4 x \epsilon^{z\mu\nu\rho} B_{\mu} \partial_{\nu} e^{z}_{\rho}.
\label{vortex_dislocation}
\end{eqnarray}

If we do not specify a restricted geometry (i.e., not just a $z$-oriented screw dislocation) and allow for the full geometric coupling, then we would find every possible permutation $\epsilon^{\lambda \mu\nu\rho} B_{\mu} \partial_{\nu} e^{\lambda}_{\rho}$. This combination of indices is unusual as it only allows ``off-diagonal" contributions of the frame fields such as those generated by screw dislocations (but not edge dislocations). This is unusual since it is not only sensitive to the Burgers' vector, but a geometric characteristic of the lattice dislocation, i.e., whether it is edge or screw. 

One interpretation of this term is that it generates a statistical interaction between an axion string (i.e., the CDW dislocation) and a geometric screw dislocation of the underlying lattice. We will illustrate this statement in two ways. We begin by setting up an axion string and a screw-dislocation as straight lines parallel to the $z$-direction. We can fix this orientation of the two strings and braid them by moving them in the $xy$-plane. We will see that this will give a Berry phase proportional to the coefficient in Eq. (\ref{vortex_dislocation}). Since the dislocation couples to the stress tensor, and the momentum currents are not quantized objects (at least for our considerations), the Berry phase here can be any irrational number. Also, since crystal dislocations are not deconfined excitations, there will naturally be other, highly non-universal, contributions to the total phase during this process.

\begin{figure}  [h!] 
\includegraphics[height=2.4in,width=2.5in]{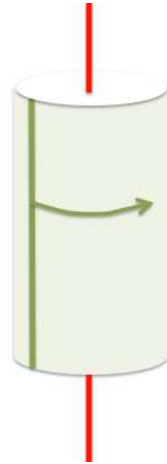}  
\caption{Statistical interaction when an axion string goes around a parallel screw dislocation line. The red line is the screw dislocation with burgers vector $\mathcal{B}_z $. The dark green line is the axion string parallel to dislocation line. The light green cylinder side surface is the trajectory of the axion string.}
\end{figure} 

As the axion string/dislocation lines are in the $z$-direction, we can treat them as an independent collection of points that braid around each other in the $xy$-plane. Hence, the Berry phase term accumulated by the winding of the screw dislocation around the axion string can be obtained by generating the analogous Hopf term. We rewrite the action as,
\begin{eqnarray}
&&\mathcal{S}=\frac{m^2 \ln(\frac{\Lambda}{m})}{4\pi^2} \int d^4x  B_\rho \partial_{\mu} e^{z}_{\nu} \epsilon^{\rho \mu \nu} \nonumber\\
&&=\int d^4 x \frac{1}{2\pi} \left[ \frac{-a^{B}_\rho \partial_{\mu} a^{e}_{\nu} \epsilon^{\rho \mu \nu} }{2\pi m^2 \ln(\Lambda/m)} +\frac{B_\rho \partial_{\mu} a^{B}_{\nu} \epsilon^{\rho \mu \nu} }{2\pi} +\frac{e^{z}_\rho \partial_{\mu}  a^{e}_{\nu}  \epsilon^{\rho \mu \nu}}{2\pi} \right] \nonumber\\
&&=\int d^4 x \frac{1}{2\pi} \left[-\frac{a^{B}_\rho \partial_{\mu} a^{e}_{\nu} \epsilon^{\rho \mu \nu} }{2\pi m^2 \ln(\Lambda/m)} +J^{B}_{\nu} a^{B}_{\nu} +\mathcal{B}_z J^{e}_{\nu} a^{e}_{\nu}\right]  \nonumber\\
&&J^{B}_{\nu}\equiv\epsilon^{\rho \mu \nu}  \partial_{\mu} B_\rho/2\pi,~  J^{e}_{\nu}\equiv\frac{\epsilon^{\rho \mu \nu}}{2\pi \mathcal{B}_z }  \partial_{\mu} e^{z}_\rho,
\end{eqnarray}\noindent  
where we have introduced auxiliary gauge fields $a^{B}$ and $a^{e}$ to decouple the mixed term, and $J^{B}, J^{e}$ are the currents of the axion string and screw dislocation line, respectively, both of which are aligned in the $z$-direction. We have treated the string currents as point-particle currents for simplicity since the strings are straight lines. We have also normalized the screw dislocation current with a factor of its Burgers vector $(\mathcal{B}_z)$ in order to have a quantized screw dislocation flux.
Now that we have this form, we can integrate out the two auxiliary gauge fields of $a^{e}$ and $a^{B}$ to obtain a Hopf term,
\begin{eqnarray}
&&\mathcal{S}_{Hopf}=\mathcal{B}_z m^2 \ln(\Lambda/m) \int d^4x ~  J^{B}_{\rho} (\frac{\epsilon^{\rho \mu \nu} \partial_{\mu} }{\partial^2}) J^{e}_{\nu}. 
\end{eqnarray}
The Hopf term evaluates to the linking number between the worldlines of each point on the two strings since they are straight-lines oriented in the $z$-direction. An effective world-line linking occurs in the $t$-$x$-$y$ spacetime and we have a (non-universal) statistical phase:
\begin{eqnarray}
\mathcal{S}_{Hopf}&&=\int dz~ \mathcal{B}_z m^2 \ln(\Lambda/m) \nonumber\\
&&= \mathcal{B}_z m^2 \ln(\Lambda/m) L_z. 
\end{eqnarray}

Using the same axion string and dislocation arrangement, we can provide a more microscopic argument for this statistical interaction between the axion string and the crystal dislocation line as well. The axion string carries a localized 1D chiral fermion propagating  along the z-direction\cite{harvey1989cosmic}. We can label each low energy mode of the chiral fermion with their momentum in the $z$-direction $k_z$. The modes remain localized on the string until they reach an energy corresponding to the bulk gap which acts as a cutoff; states beyond this energy are not localized on the string, and are not chiral. During the string braiding process, each chiral 1D $k_z$ mode traveling around the screw dislocation line will pick up a phase since the axion string is translated by the Burgers' vector of the dislocation. Each value of $k_z$ sees an effective momentum-dependent magnetic flux and picks up an Aharonov-Bohm phase. The total phase can be calculated from a sum over all the localized chiral states: 
\begin{eqnarray}
\exp\left(i\sum_{k_z} k_z \oint (e^z_j dx^j) \right)=\exp\left(i\sum_{k_z} k_z \mathcal{B}_z\right) 
\end{eqnarray}
where $\mathcal{B}_z$ is the Burgers' vector of the screw-dislocation. For a system with length $L_z$ in the $z$-direction, and with periodic boundary conditions, the $k_z$ are discretized to multiples of $\frac{2\pi}{L_z}$. Since for conventional  models the chiral modes lie inside the bulk gap, the total number of chiral states inside the gap is $m/(\frac{2\pi}{L_z})=\frac{mL_z}{2\pi}$ (recall we have set the Fermi-velocity/speed of light to unity). Therefore, $\sum_{k_z} k_z=\frac{m^2 L_z}{4\pi}$ (up to a piece which is negligible as $L_z\to\infty$). The accumulated Berry phase  is therefore
\begin{eqnarray}
\exp\left(i\sum_{k_z} k_z \oint (e^z_j dx^j)\right)
=\exp(i m^2 \mathcal{B}_z L_z).
\end{eqnarray} This matches our Hopf result up to a log correction factor. 

Apart from the mutual statistical interaction between the axion string and lattice dislocations, there is also a set of three loop statistical interactions that we can study in the axion insulator phase. First, there is a statistical interaction between two external electromagnetic flux loops threaded by an axion string. To see this we recall that the triangle diagram (Fig. \ref{fig:axialanomalydiag}) gives rise to an anomalous contribution to the conservation law for the axial current:
\begin{figure} [h!] 
\begin{center}  
\includegraphics[height=1.7in,width=2.5in]{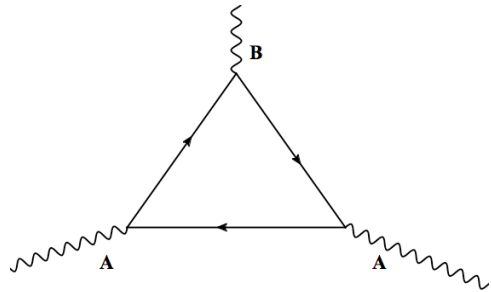}  
\caption{Conventional triangle diagram for the axial anomaly.}\label{fig:axialanomalydiag}
\end{center}  
\end{figure}  
\begin{eqnarray}
\partial_{\mu} J^{ax}_{\mu}=2m \langle \bar{\Psi} \tau_2 \Psi \rangle+\frac{1}{4\pi^2}\epsilon^{\rho \lambda \nu \mu} \partial_{\rho}A_{\lambda} \partial_{ \nu}A_{\mu}.
\end{eqnarray}
The first term on the right hand side comes from the explicit breaking of the axial/chiral symmetry at the classical level due to the non-vanishing Dirac mass, while the second term is the anomaly term from the triangle diagram. 
In the long wavelength limit,  the first term cancels the second term\cite{banerjee1999chiral}.
Hence, in the axion insulator phase, we naively find that the effect in which we are interested vanishes. However, in the presence of an external electromagnetic field, there is a path integral measure change due to the chiral transformation that gives
\begin{eqnarray}
&&\mathcal{S}_{anomalous}= \int d^4 x ~\frac{1}{4\pi^2} \epsilon^{\rho \lambda \nu \mu}  B_{\rho} A_{\lambda} \partial_{\nu} A_{\mu}. 
\label{axionbraiding}
\end{eqnarray} 
 The reader might worry that this action is not invariant under gauge transformations of $B_{\rho}$, but this is not an issue since the system has a chiral mass induced by the order parameter and the symmetry is explicitly broken.

\begin{figure}  [h!] 
\includegraphics[height=2in,width=2.5in]{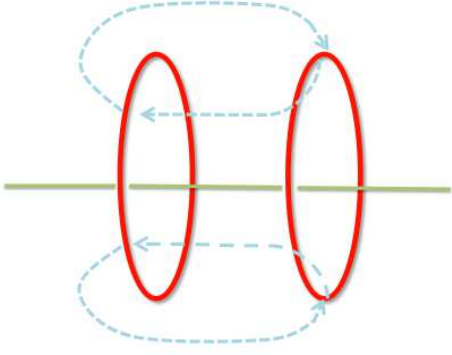}
\caption{Figure of the three loop statistical process. Red loops are  dislocation (or electromagnetic flux) lines and the green line is an axion string with $2\pi$ flux threading the other two loops. The blue dotted lines indicate the trajectory of the loop braiding between the  two dislocations.}\label{fig:threeloop}
\end{figure}
This term implies a natural three loop statistical interaction\cite{jian2014layer,wang2014braiding} between one axial flux and two electromagnetic flux loops. Imagine we have an axion string (i.e., from a vortex defect in the CDW order) going through two electromagnetic flux loops. From the coefficient of Eq. \ref{axionbraiding}, we find that  braiding one flux loop around the other (see Fig. \ref{fig:threeloop}) gives a trivial phase of $2\pi$ in this case. Although this result is not particularly exciting, we might imagine having a system of  ``fractional" Weyl fermions\cite{wang2014fractionalized} and perhaps there could exist nontrivial three loop braiding processes after gapping out the Weyl cones to form a sort of fractional axion insulator.

Finally,  we turn to the triangle diagram with one axial leg and two frame field legs (Fig. \ref{fig:torsionanomaly}).
The axial anomaly also has a contribution from this diagram at the one loop level:
\begin{figure} [h!] 
\begin{center}  
\includegraphics[height=1.5in,width=2.5in]{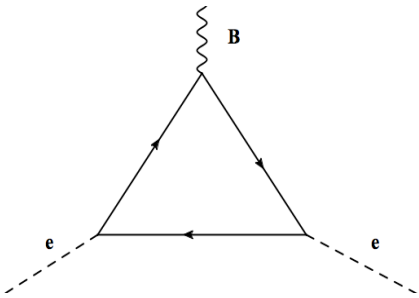}  
\caption{Frame-field contribution to axial anomaly.}\label{fig:torsionanomaly}
\end{center}  
\end{figure}  
\begin{eqnarray}
\partial_{\mu} J_{\mu}^{ax}&=& 2m\langle \bar{\Psi} \tau_2 \Psi \rangle +\frac{1}{4\pi^2}\epsilon^{\rho \lambda \nu \mu} \partial_{\rho}A_{\lambda} \partial_{ \nu}A_{\mu}\nonumber\\&+&\frac{\Lambda^2}{4\pi^2}\epsilon^{\rho \lambda \nu \mu} \partial_{\rho}e^{l}_{\lambda} \partial_{ \nu}e^{l}_{\mu}.
\end{eqnarray}
The first term in the right hand side comes from the explicit axial symmetry breaking due to the mass, while the second and third terms are the electromagnetic and geometric contributions to the anomaly term at the one loop level.  Let us only consider the geometric terms. From this anomaly we find that there is a contribution to the effective action given by
\begin{eqnarray}
\mathcal{S}_{L}= \frac{-m^2 \ln(\frac{\Lambda}{m})+\Lambda^2}{4\pi^2} \int d^4 x~\epsilon^{\rho \lambda \nu \mu}  B_{\rho} e^{l}_{\lambda} \partial_{\nu} e^{l}_{\mu} 
\label{viscosity}
\end{eqnarray} where the term proportional to $m^2$ arises from evaluating the explicit axial symmetry breaking term $ 2m\langle \bar{\Psi} \tau_2 \Psi \rangle.$ 
Just as before, the chiral transformation also changes the path integral measure of the fermion field. Keeping only the geometric contribution we find
\begin{eqnarray}
\mathcal{S}_{anomaly}= -\frac{\Lambda^2}{4\pi^2}\int d^4 x ~ \epsilon^{\rho \lambda \nu \mu}  B_{\rho} e^{l}_{\lambda}\partial_{\nu} e^{l}_{\mu}. 
\end{eqnarray}
Thus, in total we have the effective theory for the geometric response
\begin{eqnarray}
&&\mathcal{L}_{B,e}= -\frac{m^2 \ln(\frac{\Lambda}{m})}{4\pi^2} \epsilon^{\rho \lambda \nu \mu}  B_{\rho} e^{l}_{\lambda} \partial_{\nu} e^{l}_{\mu}.
\label{threeloop}
\end{eqnarray}

This term also can be interpreted as a Berry phase accumulated by a three loop statistical interaction.
Imagine we have an axion string threading two dislocation loops. Then  Eq. \ref{threeloop} determines the phase of a process where we wind one dislocation loop around the other as in Fig. \ref{fig:threeloop}, and $2\pi m^2 \ln(\frac{\Lambda}{m})$ is the phase it accumulates.

\section{Two Routes to the Chiral Magnetic Effect and Analogous Geometric Effects}
In our discussions so far we have focused primarily on geometric response phenomena and statistical interactions between axion strings and lattice dislocations. However, in this section we provide two possible routes to an electromagnetic response property. 
The first one is due to the cross-coupling between the external electromagnetic field and the geometrical distortions in a gapless WSM phase, while the other is due to the fluctuations of the CDW order responsible for gapping a WSM to form the axion insulator. For the former response we essentially  show that having a non-vanishing screw dislocation density can generate a nonzero axial potential $B_0$ which shifts the nodal energies of the two Weyl cones differently. It is well-known\cite{zyuzin2012weyl,son2012berry,stephanov2012chiral,vazifeh2013electromagnetic,yamamoto2015generalized,huang2015observation,kharzeev2014chiral,li2014observation,son2013chiral}, though a bit subtle and controversial\cite{chang2015chiral,sumiyoshi2015torsional,chang2015chiralm}, that a non-vanishing $B_0$ will give rise to a chiral magnetic effect. Hence a dislocation density will also produce such an effect in the presence of a magnetic field. For the latter response, we will argue that the usual electromagnetic response of a WSM can survive into the gapped phase because of the space-time dependence of the CDW order parameter. We finish this section with a discussion of some analogous geometric effects including a Hall viscosity response
of the axion insulator phase, a chiral geometric effect, and a geometric Witten effect. 

\subsection{Electromagnetic Effects}
Let us start with a phenomenon in the gapless WSM. The effect of interest is the result of a mixed coupling between electromagnetic and geometric fields, and is effectively  a dislocation-induced chiral magnetic effect.
In a WSM, an axial chemical potential (relative energy shift of the nodes) would result in a CME where a non-vanishing magnetic flux in the $i$-direction leads to charge current in the $i$-direction via the effective response action: 
\begin{eqnarray}
&&\mathcal{L}_{CME}=\frac{1}{4\pi^2}b_0 \epsilon^{ijk} A_i \partial_j A_k\nonumber\\
&&\vec{J}=\frac{b_0\vec{B}^{EM}}{2\pi^2}. \label{eq:LforCME}
\end{eqnarray}
We have already seen that in an axion insulator the screw dislocation density couples with the axion gauge potential $B_0$ (see Eq. \ref{vortex_dislocation}). If we generate a non-vanishing dislocation density, we then expect an effective axial potential will be generated, and a similar chiral magnetic effect will be produced.

\begin{figure}  [h!] 
\includegraphics[height=1.5in,width=2.7in]{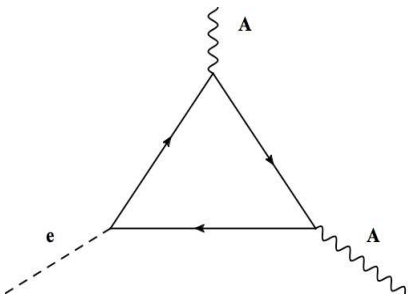}  
\caption{CME from dislocation density.}\label{fig:CMEdislocation}
\end{figure}

To explicitly demonstrate this effect we calculate the triangle diagram in Fig. \ref{fig:CMEdislocation}.
For simplicity, here we assume that the screw dislocation line is parallel to the $z$-axis, but this also applies to any other direction. After a straightforward calculation (see Appendix A), we arrive at the effective theory
\begin{eqnarray}
\mathcal{L}_{CME}=\frac{1}{4\pi^2} \epsilon^{ij}  \partial_{j} e^{z}_{i} \epsilon^{\mu \nu \rho} A_{\mu} (\partial_{\nu} A_{\rho}).  \label{eq:LforCME2}
\end{eqnarray}
By direct comparison to Eq. \ref{eq:LforCME} we see that the dislocation density $\epsilon^{ijk}  \partial_{j} e^{k}_{i}$ plays an analogous role to the axial potential. Interestingly, this is also one of the few geometric terms where the coefficient is independent of the high-energy cutoff.

We can provide a microscopic interpretation of this result as well. Before we go into the dislocation induced CME, we first review the microscopic, continuum picture for the CME generated by an axial chemical potential.
We begin by assuming that we have an external, uniform magnetic field $B_z$ in the $z$-direction. The energy spectrum in the $xy$-plane will form Landau levels. The CME is determined by the  zeroth Landau level, for which the energy dispersion is $E(k_z)=\pm k_z$ (we have chosen a velocity $|v_F|=1$). The degeneracy for each $k_z$ is $\frac{B_z}{2\pi} L^2$ where $L^2$ is the area of the $xy$-plane. Now let us recall a standard continuum argument for the CME (although if we wanted to be precise we should consider lattice effects as well as a possible low-frequency magnetic field at a non-static level\cite{zhang2015quantum,bulmash2015quantum,huang2015observation,kharzeev2014chiral,li2014observation,son2013chiral}). We truncate the Brillouin zone to the range $-\Lambda$ to $\Lambda$, and then we turn on an axial chemical potential $b_0$ which shifts the energies of the left/right Fermi points. The induced current density is
\begin{eqnarray}
&&J_z=\frac{1}{2\pi}\left(\frac{B_z}{2\pi}\right) \int_{k_z \in occ.} dk_z \frac{\partial E}{\partial k_z}\nonumber\\
&&=\frac{ b_0 B_z}{2\pi^2}
\end{eqnarray}
which matches the CME response term in the effective theory in Eq. \ref{eq:LforCME}.

Now we can apply a similar argument to see the dislocation induced CME. Let us assume translation invariance in the $z$-direction and fix the momentum $k_z.$ In addition to the external magnetic field $B_z$, assume a uniform dislocation density (for simplicity) with Burgers vector $\mathcal{B}_z.$ This acts as an effective magnetic flux in the $z$-direction, but couples to momentum charge as $k_z \mathcal{B}_z$ at the fixed value of $k_z$. Therefore, for each $k_z$, we have Landau levels with degeneracy $\frac{1}{2\pi}(k_z \mathcal{B}_z +B_z)L^2$. In the zeroth Landau level, there exists a gapless chiral mode lying inside the Landau level gap $\Delta \sim 2\sqrt{2B_z}$ (where we have set the Fermi velocity equal to unity).
The induced current density is
\begin{eqnarray}
J_z&&= \frac{1}{4\pi^2}\int dk_z  (k_z \mathcal{B}_z+B_z)  \frac{\partial E}{\partial k_z}\nonumber\\
&&=\frac{1}{4\pi^2}\int_0^{\sqrt{2B_z}} dk_z (k_z \mathcal{B}_z+B_z) \nonumber\\ 
&&~~~~~-\frac{1}{4\pi^2}\int^{0}_{-\sqrt{2B_z}} dk_z (k_z \mathcal{B}_z+B_z)\nonumber\\
&&=\frac{1}{2\pi^2} \mathcal{B}_z B_z .
\end{eqnarray} which agrees with our effective theory.

Now let us discuss the second effect mentioned above by describing the electromagnetic response of the axion insulator due to the fluctuating axion field. The axion insulator phase is gapped since the Weyl nodes are eliminated by the CDW order. Thus, there is naively no expectation that anything like an anomalous Hall effect or chiral magnetic effect should persist in the insulator phase; especially since the usual definition of $b_\mu$ is no longer well-defined without Weyl nodes in the band structure. However, 
as the mass term of the insulator is generated by the CDW order parameter with wave vector $2\bm{b}$, the mass is actually spatially dependent $\vec{m}=m \exp(i 2 b_i r_i)$ and \emph{remembers} the nodal configuration of the Weyl nodes from the parent WSM state. When the CDW order is free of string defects, we have $B_{\mu}=b_{\mu}$ and the theory in Eq. \ref{axionbraiding} then becomes,
\begin{eqnarray}
\mathcal{L}= \frac{1}{4\pi^2} \epsilon^{\rho \lambda \nu \mu}  b_{\rho} A_{\lambda} \partial_{\nu} A_{\mu}.
\end{eqnarray}
This describes the response in the axion insulator phase, and it is similar to the usual anomalous Hall effect/chiral magnetic effect response for a Weyl semimetal. However, the responses in the axion insulator arise from the space and time dependence of the CDW order parameter, not from the momentum and energy separation of gapless Weyl cones. Interestingly, if the Weyl nodes were separated in energy, the CDW would have to have a built-in time dependence to precisely nest the energy-separated nodes, and this would leave some memory of the chiral magnetic response. 

Thus, once a Weyl semimetal is gapped to form an axion insulator, a time dependent CDW order parameter can generate the CME, and spatial dependence will produce an anomalous Hall effect. Therefore, we also expect there to be an anomalous Hall effect and CME in the axion insulator, with the same coefficient as in the parent, gapless WSM phase.

\subsection{Geometric Effects}
In the axion insulator phase we recall the effective action showing the coupling between axion strings and crystal dislocations from Eq. (\ref{threeloop}): 
\begin{eqnarray}
\mathcal{L}= \frac{m^2 \ln(\frac{\Lambda}{m})}{4\pi^2} \epsilon^{\rho \lambda \nu \mu}  B_{\rho} e^{l}_{\lambda} \partial_{\nu} e^{l}_{\mu}.\nonumber 
\end{eqnarray}
Hence, the geometric version of the anomalous Hall effect  exists if we have a spatial-dependent CDW order parameter $\vec{m}=|m| \exp(i 2b'_z z)$ which produces the term in the stress response \begin{eqnarray}
T^{l}_{i} = \frac{m^2 \ln(\frac{\Lambda}{m}) }{2\pi^2} \epsilon^{i j}  b'_{z} \partial_{0} e^{l}_{j}. 
\end{eqnarray}
Here $i, j$ run over $x,y$. This term represents  the Hall viscosity in 3D where a shear deformation would result in a momentum current perpendicular to the strain rate. Since a generic axion insulator naturally has a spatially-dependent order parameter (due to the $-2{\bf{b}}\cdot{\bf{r}}$ background contribution to the axion field) this result implies that they will generically have a Hall viscosity. 

Furthermore, if we have a  time-dependent CDW order parameter $\vec{m}=|m| \exp(i 2b'_0 t)$, we expect a geometric version of the chiral magnetic effect,
\begin{eqnarray}
T^{l}_{i} = \frac{m^2 \ln(\frac{\Lambda}{m})}{2\pi^2} \epsilon^{0 ijk}  b'_{0} \partial_{j} e^{l}_{k} 
\end{eqnarray} which implies that if we have a finite dislocation density $\partial_{j} e^{l}_{k}$ in the $jk$-plane (with Burgers vector in the $l$-direction), we would obtain a momentum current response $T^{l}_{i}$ in the direction perpendicular to the $jk$-plane with momentum pointing in the $l$-direction.

Finally, we can rewrite Eq. (\ref{threeloop}) as \begin{eqnarray}
\mathcal{L}= \frac{m^2 ln(\frac{\Lambda}{m})}{8\pi^2}  \theta \epsilon^{\rho \lambda \nu \mu}  \partial_{\rho} e^{l}_{\lambda} \partial_{\nu} e^{l}_{\mu} \label{eq:geomthetaterm}
\end{eqnarray}\noindent assuming the CDW order induces a mass $\vec{m}=m \cos \theta+i m\tau_2 \sin \theta,$ and
$\theta$ is the phase angle of the CDW order parameter.  This term is in analogy with the electromagnetic theta term $\propto \theta \epsilon^{\rho \lambda \nu \mu}  \partial_{\rho} A_{\lambda} \partial_{\nu} A_{\mu}$ that also appears in an axion insulator.
It is known that the usual electromagnetic theta term gives rise to the Witten effect, so let us consider an analogous effect from Eq. \ref{eq:geomthetaterm}.
The stress-energy tensor receives one contribution of the form
\begin{eqnarray}
&&T^{l}_{\lambda} =\frac{m^2 \ln(\frac{\Lambda}{m})}{4\pi^2}  \theta \epsilon^{\lambda \rho \nu \mu}  \partial_{\rho} \partial_{\nu} e^{l}_{\mu} 
\label{monopole}
\end{eqnarray}\noindent from this term in the effective action. 
Akin to the magnetic monopole configuration of the electromagnetic field, we can use $\frac{1}{2}\epsilon^{0 \rho \nu \mu}  \partial_{\rho} \partial_{\nu} e^{l}_{\mu}$  as a definition of a geometric monopole $\rho_m^{l}$ carrying dislocation/Burgers' vector flux as discussed above. In analogy to the conventional Witten effect in an axion insulator, where a magnetic charge binds electric charge, in this case a momentum density $T^{l}_{0}$ is bound to a dislocation monopole charge $\rho_m^{l},$ and a momentum current carries a dislocation monopole current. Hence, this is a purely geometric Witten effect.

\section{Conclusion}
In summary, we have explored the  properties of Weyl semimetals and axion insulators coupled to geometry. The novel interplay between axion strings and lattice dislocation defects were studied through both the effective field theory and the microscopic arguments.

Unlike in a time-reversal invariant topological insulator, the axion field in an axion insulator is dynamical and can generate string/vortex defects in configurations of the CDW order parameter.  The linear coupling between the axion string and the lattice geometry induces a statistical interaction between the axion strings and screw dislocation lines. The cubic coupling between the axion string and geometry fields is a signature of a three-loop statistical interaction between two dislocation loops penetrated by an axion string. In addition, we also derived an effective response for a chiral magnetic effect induced by screw dislocation density instead of an axial chemical potential, i.e., an energy imbalance between Weyl cones. This  provides another possible way to measure a nontrivial electromagnetic response in Weyl semi-metals. We also presented several purely geometric response effects that might also be measured in future experiments. 

\begin{acknowledgments}
We are grateful to Y. Lu, J. Teo, and Z. Wang for insightful comments and discussions. This work was supported in part by the National Science Foundation through grants DMR-1408713 (YY) at the University of Illinois, grants PHY11-25915(YY) at The Kavli Institute for Theoretical Physics, the Brain Korea 21 PLUS Project of Korea Government (GYC). YY thank the hospitality of IASTU where the idea of this work initiated. TLH. is supported by the US National Science Foundation under grant DMR 1351895-CAR.
\end{acknowledgments}

\appendix

\section{Appendixes}
\subsection{Detailed calculation of the CME induced by dislocation}

For the chiral magnetic effect induced by screw dislocations, we consider the triangle diagram in Fig. \ref{fig:CMEdislocation}.
\begin{eqnarray}
&&\frac{\delta Z}{\delta e^{z}_{\rho} \delta A_{\mu} \delta A_{\nu} }=-i\langle J_{\mu} J_{\nu}  T^{z}_{\rho} \rangle\equiv V^{z}_{ \mu \nu \rho }.
\label{tri}
\end{eqnarray}
Here we assume the screw dislocation line is along the $z$-axis, but this derivation can be easily generalized to other directions.
Instead of calculating the triangle diagram directly, we study the anomaly of the charge current in the presence of dislocations. The non-conserved part(anomalous contribution) of this polarization tensor contributes to the CME response. To test the charge conservation of the polarization tensor in Eq.~\ref{tri}, we calculate
\begin{widetext}
\begin{eqnarray}
&&q_{\mu} V^{z}_{ \mu \nu  \rho}(q,p,(-q-p))+p_{\mu} V^{z}_{ \nu \mu \rho}(q,p,(-q-p)) \nonumber\\
&&=\int d\vec{k} d\omega~ [q_{\mu} G(k) \gamma_{\mu} G(k+q) \gamma_{\nu} G(k+q+p) \gamma_{\rho}(2k+p+q)
+p_{\mu}G(k) \gamma_{\nu} G(k+q) \gamma_{\mu} G(k+q+p) \gamma_{\rho}(2k+p+q)]. \nonumber
\end{eqnarray} 
Here $G(\bm{p})=(p_0 \tau_1+p_i \tau_2 \otimes \sigma_i)^{-1}$ is the Green function of a single Weyl cone. Recall the identity that 
\begin{eqnarray}
q_{\mu} \gamma_{\mu} =G^{-1}(k+q)-G^{-1}(k).
\end{eqnarray} 
We can rewrite Eq. \ref{tri} as 
\begin{eqnarray}
&&=\int d\vec{k} d\omega~ [q_{\mu} G(k) \gamma_{\mu} G(k+q) \gamma_{\nu} G(k+q+p) \gamma_{\rho}(2k+p+q)_z
+p_{\mu}G(k) \gamma_{\nu} G(k+q) \gamma_{\mu} G(k+q+p) \gamma_{\rho}(2k+p+q)_z]  \nonumber\\
&&=\int d\vec{k} d\omega~[ G(k) \gamma_{\nu} G(k+q+p) \gamma_{\rho}(2k+p+q)_z- G(k+q) \gamma_{\nu} G(k+q+p) \gamma_{\rho}(2k+p+q)_z\nonumber\\
&&+G(k) \gamma_{\nu} G(k+p) \gamma_{\rho}(2k+p+q)_z-G(k) \gamma_{\nu} G(k+q+p) \gamma_{\rho}(2k+p+q)_z]\nonumber\\
&&=\int d\vec{k} d\omega~ G(k) \gamma_{\nu} G(k+p) \gamma_{\rho}(2k+p+q)_z- G(k+q) \gamma_{\nu} G(k+q+p) \gamma_{\rho}(2k+p+q)_z\nonumber\\
&&=\int d\vec{k} d\omega~ [G(k) \gamma_{\nu} G(k+p) \gamma_{\rho}(2k+p+q)_z- G(k+q) \gamma_{\nu} G(k+q+p) \gamma_{\rho}(2k+p+3q)_z+G(k+q) \gamma_{\nu} G(k+q+p) \gamma_{\rho}(2q)_z]. \nonumber\\
\end{eqnarray}
The first two terms are related by a shift of $q$ in the integral, but both of them are linearly divergent so they do not simply cancel. 
\begin{eqnarray}
&&\int d\vec{k} d\omega~ [G(k) \gamma_{\nu} G(k+p) \gamma_{\rho}(2k+p+q)_z- G(k+q) \gamma_{\nu} G(k+q+p) \gamma_{\rho}(2k+p+3q)_z+G(k+q) \gamma_{\nu} G(k+q+p) \gamma_{\rho}(2q)_z]\nonumber\\
&&=\int d\vec{k} d\omega~  q_i  \partial_{k_i} [-G(k-p-q) \gamma_{\nu} G(k-q) \gamma_{\rho}(2k)_z]+G(k+q) \gamma_{\nu} G(k+q+p) \gamma_{\rho}(2q)_z\nonumber\\
&&=\int d\vec{k} d\omega~ q_i  \partial_{k_i} [-G(k-p-q) \gamma_{\nu} G(k-q) \gamma_{\rho}]2k_z.
\end{eqnarray}
Thus, we have
\begin{eqnarray}
&&V^{z}_{ \mu \nu  \rho}(q,p,(-q-p))=\int d\vec{k} d\omega~ q_i k_z [\partial_{k_i}[G(k) \gamma_{\mu} G(k+q) \gamma_{\nu} G(k+q+p) \gamma_{\rho}].
\end{eqnarray} 
For $i\neq z$, we have
\begin{eqnarray}
&&V^{z}_{ \mu \nu  \rho}(q,p,(-q-p))=  q_i  \int d\vec{k} d\omega~ \partial_{k_i}[G(k) \gamma_{\mu} G(k+q) \gamma_{\nu} G(k+q+p) \gamma_{\rho} k_z]\nonumber\\
&&=q_i \lim_{k\rightarrow \infty } k^2 k_i Tr[\gamma_z \gamma_{\mu} \gamma_i \gamma_{\nu} \gamma_j \gamma_{\rho}] (q+p)_j \frac{k_i k_z^2}{(k^2)^3}\nonumber\\
=&& q_i (q+p)_j  \epsilon^{ \mu i \nu}\epsilon^{ j \rho z}.
\end{eqnarray} 
For $i=z$, we have
\begin{eqnarray}
&&V^{z}_{ \mu \nu  \rho}(q,p,(-q-p))=\int d\vec{k} d\omega~ q_z k_z [\partial_{k_z}[G(k) \gamma_{\mu} G(k+q) \gamma_{\nu} G(k+q+p) \gamma_{\rho}]\nonumber\\
=&&\int d\vec{k} d\omega~ (-q_z) [G(k) \gamma_{\mu} G(k+q) \gamma_{\nu} G(k+q+p) \gamma_{\rho}]
+ q_z  [\partial_{k_z}[G(k) \gamma_{\mu} G(k+q) \gamma_{\nu} G(k+q+p) \gamma_{\rho} k_z]\nonumber\\
=&& q_z (q+p)_j  \epsilon^{ \mu z \nu}\epsilon^{ j \rho z}.
\end{eqnarray} 

This gives the result 
\begin{eqnarray}
\partial_{l} J_{l}=\frac{1}{2\pi^2} \epsilon^{ij} \epsilon^{\mu \nu \rho} (\partial_{\mu} A_{\rho}) \partial_{\nu} \partial_{j} e^{z}_{i}.
\end{eqnarray}
Here $ij$ runs over $xy$, and $\mu \nu \rho$ runs over $xyz$.
Thus one has
\begin{eqnarray}
\mathcal{L}_{CME}=\frac{1}{4\pi^2} \epsilon^{ij}  \partial_{j} e^{z}_{i} \epsilon^{\mu \nu \rho} A_{\mu} (\partial_{\nu} A_{\rho}).  
\end{eqnarray}

\end{widetext}



\providecommand{\noopsort}[1]{}\providecommand{\singleletter}[1]{#1}%

\end{document}